\documentclass[prl,nofootinbib,11pt]{revtex4}

\usepackage{amsmath,amssymb}
\usepackage{color,graphicx, slashed}
 
\newcommand{\be}{\begin{equation}}
\newcommand{\ee}{\end{equation}}

\renewcommand{\Im}{\text{Im\,}}

\newcommand{\ren}{\text{ren}}

\begin{document}

\title{One-Loop Correction to the Higgs Mass}
\author{Kang-Sin Choi}
\email{kangsin@ewha.ac.kr}
\affiliation{Scranton Honors Program, Ewha Womans University, Seoul 03760, Korea} 
\affiliation{Institute of Mathematical Sciences, Ewha Womans University, Seoul 03760, Korea}

\begin{abstract}
We present the complete one-loop corrected Higgs mass within the Standard Model. It is crucial to identify the observable mass as the renormalized one that depends on the external momentum. The correction is finite and dominated by the loops involving the $W$ bosons and the top quark. These contributions exhibit a power-law running with momentum, which may be verifiable at the Large Hadron Collider.
\end{abstract}

\maketitle

We calculate the complete one-loop corrections to the Higgs mass squared, which is the major missing element in understanding the quantum nature of the Standard Model. The calculation of correction amplitude to the mass squared, or the self-energy, has been explored ever since the work of Veltman \cite{Veltman:1980mj} (see also \cite{Fleischer:1980ub, Martin:2007pg, Grange:2010zz, Holthausen:2011aa, Lynn:2011aa, Bezrukov:2012sa, Vieira:2012ex, Farina:2013mla}). However, these attempts have not successfully yielded a meaningful quantity because the self-energy has been considered quadratically divergent. 

Nevertheless, faithful renormalization yields a finite physical observable that exhibits scale dependence. For instance, the top quark contribution has been obtained as finite and scale-dependent in \cite{Choi:2023cqs}.
Since the scalar mass squared is (super-)renormalizable, the divergent part can be systematically cancelled in the renormalized mass, if we keep the counterterms \cite{Hepp:1966eg}. 
It is essential to identify the renormalized mass with the full counterterms. 

Further evidence for the finiteness comes from the optical theorem: the imaginary parts of the one-loop amplitudes are equal to two-body decay widths, which are known to be finite \cite{Ellis:1975hr, Lee:1977eg, Gunion:1989we}. Since divergences in the self-energy do not separate into real and imaginary parts, the entire amplitude must be finite.

We note that we can only indirectly observe the mass through the propagator $i/\Gamma^{(2)}(p^2)$ in the $S$-matrix. For the unrenormalized Higgs $h_{\text{un}}$, the two-point vertex in the interacting vacuum is
\be \label{twopoint}
  \Gamma^{(2)}(p^2) \equiv -i \langle h_{\text{un}} (-p) h_{\text{un}} (p) \rangle_{\text{1PI}}  = p^2-m_B^2- \tilde \Sigma(p^2), 
\ee
where $m_B^2$ is the bare mass squared in the defining Lagrangian.
The self-energy $\tilde \Sigma(p^2)$ is characterized by the sum of one-particle-irreducible (1PI) amplitudes with two external Higgs fields $h$. 
We define the pole mass $m_h^2$ satisfying
\be \label{polemass}
 \Gamma^{(2)}(m_h^2) = m_h^2 - m_B^2 - \tilde \Sigma(m_h^2) \equiv 0,
\ee
regarded as the renormalization condition in the physical scheme. This requires no fine-tuning: we determine the unknown $m_B^2$ from the remaining terms in Eq.  (\ref{polemass}), which depend on $m_h^2$ \cite{Coleman:2018mew}. In other words, we may expand the theory using the pole mass $m_h^2$ instead of the bare mass. Through field-strength renormalization $h(x) \equiv  ( 1- \frac{ d \tilde \Sigma}{d p^2}(m_h^2))^{1/2} h_{\text{un}}(x)$, the dressed field effectively behaves as if it has approximate mass $m_h$ with unit residue in the propagator \cite{Coleman:2018mew, Cheng:2000ct, Hollik:1988ii, Choi:2024cbs}
\be \label{gamma2}
\begin{split}
 \Gamma^{(2)}_{\ren}(p^2) & \equiv -i \langle h (-p) h(p) \rangle_{\text{1PI}}\\
   &= p^2 - m_h^2 - \tilde \Sigma(p^2) + \tilde \Sigma(m_h^2) + (p^2 -m_h^2) \frac{d\tilde \Sigma}{d p^2} (m_h^2) + {\cal O}\big((p^2-m_h^2)^2\big).
\end{split}
\ee 
In the scattering experiment, we are only able to indirectly observe the entire combination \eqref{gamma2}, since only the fully interfered amplitude appears in the $S$-matrix; by any physical detection process, no single term is separately observable. Accounting for quantum corrections and comparing with the free propagator, we can extract the {\em renormalized mass} \cite{Choi:2023cqs, Choi:2024cbs}
\be \label{SlidingMass} 
\begin{split}
 m^2(p^2) &\equiv p^2 - \Gamma^{(2)}_{\text{ren}}(p^2) \\
 &= m_h^2 + \tilde \Sigma(p^2) - \tilde \Sigma(m_h^2) - (p^2 -m_h^2) \frac{d\tilde \Sigma}{d p^2} (m_h^2) .
\end{split}
\ee
Since this mass is expressed as the difference of the same function at different points, it is insensitive to the high-momentum contribution in the loop. The quadratic divergence of $\tilde \Sigma(p^2)$ is cancelled by that in $\tilde \Sigma(m_h^2)$ and the logarithmic divergence is cancelled by the next-order term in $(p^2 -m_h^2)$ \cite{Bogoliubov:1957gp,Hepp:1966eg, Zimmermann:1967, Zimmermann:1969jj, Choi:2023cqs, Choi:2024cbs}. The finite quantity is unambiguous because we do not artificially remove divergences.

The terms in the second line, except $m_h^2 + \tilde \Sigma(p^2)$, are conventional counterterm parameters in the momentum space, which we should keep faithfully to satisfy the renormalization condition (\ref{polemass}) and the unit residue condition.\footnote{Using the counterterms ${\cal L} \supset \frac12 \partial_\mu h \partial^\mu h + \frac12 \delta_Z \partial_\mu h \partial^\mu h - \frac{1}{2} m_h^2 h^2 - \frac12 \delta_m h^2$, we may also define the renormalized self-energy as $\tilde \Sigma_{\ren}(p^2) \equiv \tilde \Sigma(p^2) - p^2 \delta_Z +\delta_m,$ the renormalization conditions $\tilde \Sigma_{\ren} (m_h^2)=0, \frac{d \tilde \Sigma_{\ren}}{dp^2} (m_h^2) = 0$ gives the same result \eqref{gamma2}. See, e.g. \cite{Peskin:1995ev} (10.32).} The counterterms not only remove the divergences but also contribute to the physical quantity. We can readily verify that any regularization yields the same result, as the renormalized mass is finite \cite{Zimmermann:1969jj, Choi:2024cbs}.	

It follows that {\em the momentum dependence is the observational consequence} of the loop corrections. We measure the mass $m^2 (p^2)$ at different energy scales specified by the {\em external momentum $p^2$} from the reference mass $m^2$ \cite{Choi:2023cqs, Choi:2024cbs}. This external momentum here removes the usual renormalization scale $\mu^2$.

\begin{figure}
\begin{center}
\includegraphics[scale=0.5]{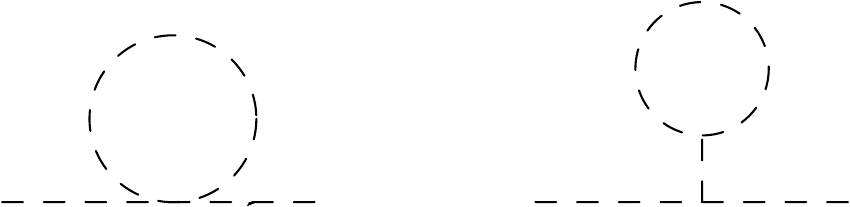}
\end{center}
\caption{Possible but vanishing `tadpole' loops. We displayed the diagrams with the Higgs in the loop; however, we have similar ones with fermions, vector bosons and Goldstone bosons. \label{Figvanishing}}
\end{figure}

\begin{figure}
\begin{center}
\includegraphics[scale=0.57]{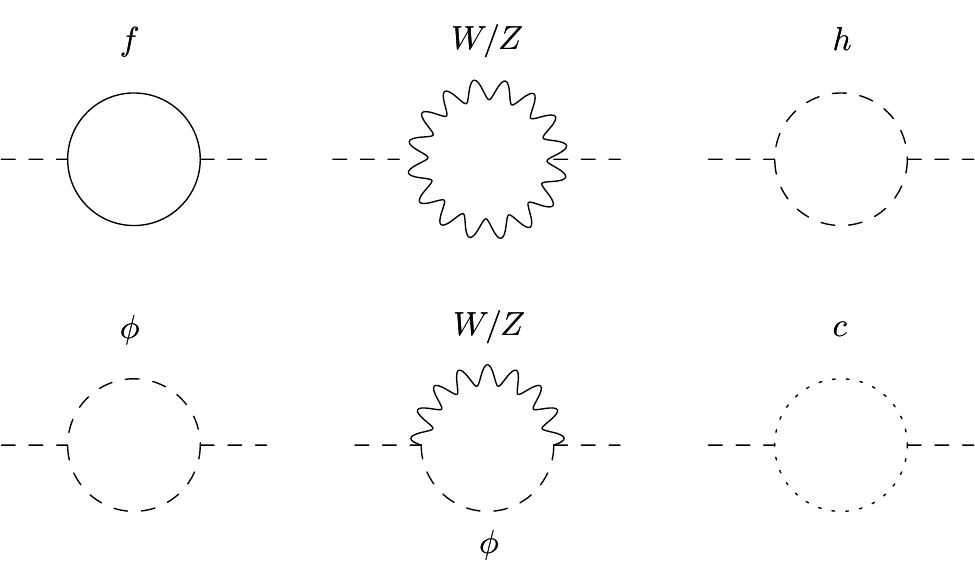}
\end{center}
\caption{Nonvanishing `bubble' loops in the Feynman--'t Hooft gauge involving the fermions $f$, the Higgs boson $h$, the $W^\pm$ bosons and the $Z$ boson. The Goldstone bosons $\phi^+, \phi^- , \varphi$ are equivalent to the longitudinal components of the massive $W^+, W^-, Z$, respectively. We also have Fadeev--Popov ghosts $c,c^+$ and $c^-$. \label{figloops}}
\end{figure}

An immediate consequence is that amplitudes of the `tadpole' diagrams, displayed in Fig. \ref{Figvanishing}, that are independent of the external momentum $p^2$, vanish trivially \cite{Choi:2024cbs}. Without this cancellation, we would suffer quadratically divergent loops of the $W$ and the $Z$ bosons of quartic couplings. This removes most of the quadratic scaling amplitudes with dimensionless couplings.
The rest, bosonic `bubble' types with dimensionful coupling, shown in Fig. \ref{figloops}, give logarithmic contributions, except the ones with top quark and longitudinal vector bosons.

The relevant interaction Lagrangian is obtained by the electroweak symmetry breaking
\be \begin{split}
{\cal L} \supset & -\frac12 m_h^2 h^2 \Big(1+ \frac{h}{2v} \Big)^2 + \left[ m_W^2 W^{+\mu } W_{\mu}^- + \frac12 m_Z^2 Z_\mu Z_\mu \right] \left(1 + \frac{h}{v} \right)^2 \\
 & - \sum_f \frac{m_f}{v} h \overline f f  -  \frac{m_Z^2}{v } h \bar c c -  \frac{ m_W^2}{v} h (\bar c^+  c^- + \bar c^- c^+) .
 \end{split}
\ee
Here, $m_h \simeq 125, m_W \simeq  80, m_Z \simeq 91, v\simeq 246$, all in GeV, are the Higgs pole mass, the $W$ boson mass, the $Z$ boson mass and the vacuum expectation value of electroweak symmetry breaking, respectively. It is well-known that spontaneously broken gauge theories remain renormalizable with the same counterterms as unbroken theories. We omit the counterterms. Also, the barred and unbarred Fadeev--Popov ghosts $\bar c,  c,\bar c^+,  c^-, \bar c^- ,c^+$ are independent.
In the Feynman--'t Hooft gauge that we employ, we have additional Goldstone boson interactions
\begin{equation} \begin{split}
 {\cal L} \supset & + \frac{m_W}{v} W^{- \mu}  h i \partial_\mu \phi^+   - \frac{m_W}{v} W^{- \mu}i  \phi^+ \partial_\mu h    - \frac{m_W}{v} W^{+\mu} h i \partial_\mu \phi^- + \frac{m_W}{v} W^{+\mu} i \phi^- \partial_\mu  h  \\
& + \frac{m_Z}{v}  Z^\mu  h \partial_\mu \varphi - \frac{m_Z}{v}  Z^\mu \varphi \partial_\mu  h    - \frac{m_h^2}{v} \phi^+ \phi^- h - \frac{m_h^2}{2v} \varphi^2 h - m_W^2 \phi^+ \phi^- -\frac12 m_Z^2 \varphi^2.
\end{split}
\end{equation}
The Goldstone bosons $\phi^+,\phi^- \equiv (\phi^+)^\dagger,\varphi$ are equivalent to the longitudinal components of the massive bosons $W^+, W^-, Z$, respectively.

\begin{figure}[t]
\includegraphics[scale=0.5]{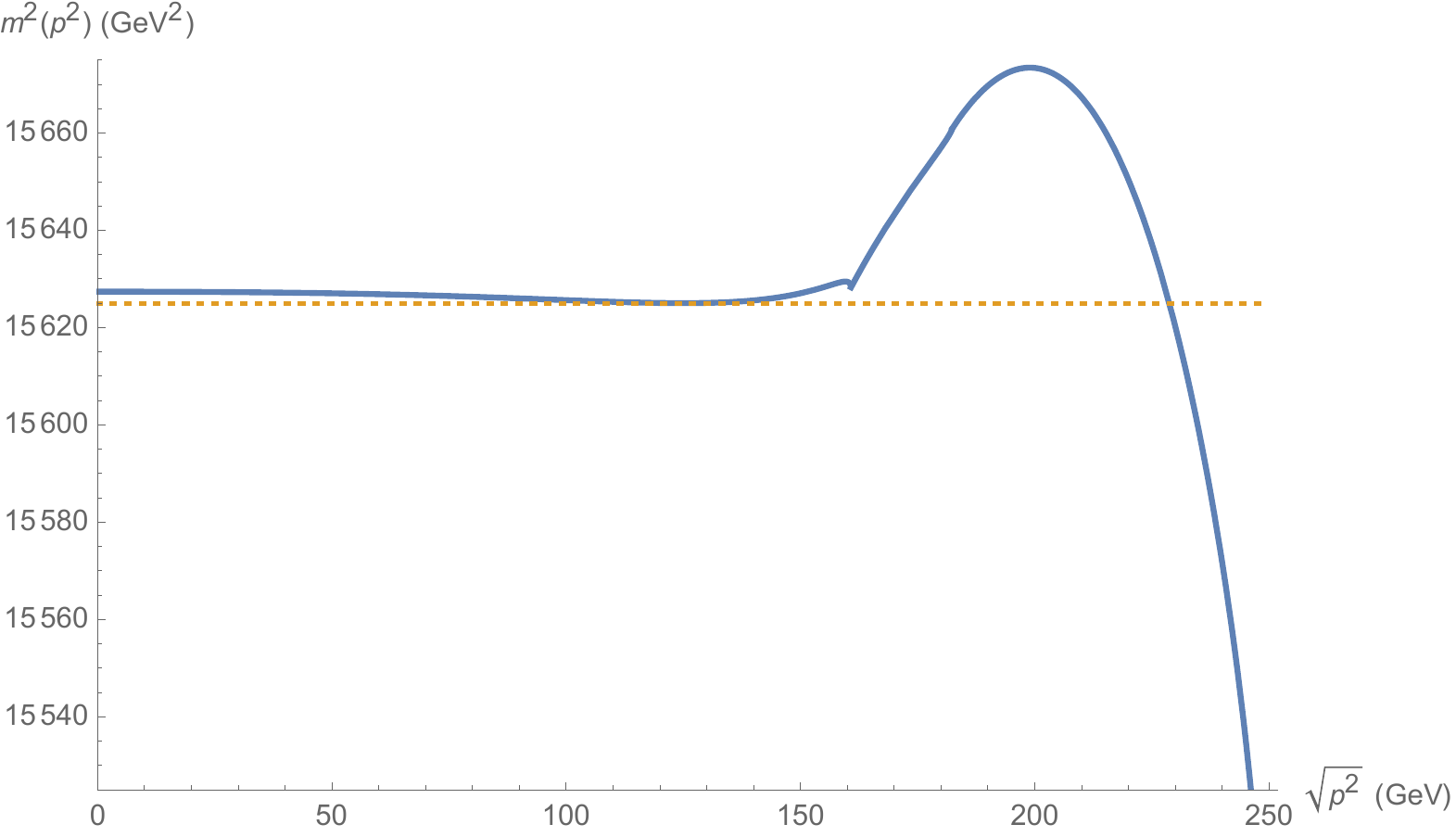} 
\caption{One-loop corrected Higgs mass squared as a function of the external momentum $\sqrt{p^2}$, from \eqref{SlidingMass} and \eqref{totalSigma}. It is characterized by a $0.31\%$ deviation peak at $199\,$GeV and, beyond that, drastically decreasing behavior dominated by the top quark contribution. We set the pole mass squared $m_h^2= 125^2 =15625 \,\text{(GeV}^2),$ indicated by the dotted line. At $\sqrt{p^2} = 2 m_Z$ and $\sqrt{p^2} = 2  m_W$, the curve exhibits cusps, indicating the opening of decay channels. We used $v=246\,\text{GeV}$ and $m_t=172.76 \,\text{GeV}$. \label{figMain}}
\end{figure}

We compute the Higgs self-energy $\tilde \Sigma (p^2)$ at one loop as
\be \label{totalSigma}
 \tilde \Sigma(p^2) = \sum_{a=W,Z,h,t} \tilde \Sigma_{1}^a(p^2) + {\cal O}(\hbar^2),
\ee
where the subscript indicates the one-loop. Individual contributions can be computed using the standard method employing the Feynman rules. We express the loop integrals in terms of bases
\begin{align}
 Q_{2n} (p,m) &= i \int\frac{d^4 k}{(2\pi)^4} \frac{1}{(k^2- \Delta_m(p^2))^{2-n}},\quad \Delta_m(p^2) = m^2 - x(1-x)p^2, \\
 R_{2n} (p,m) &= \int_0^1 dx\, Q_{2n}(p,m) .
\end{align}
Here, $\Delta_m(p^2)$ is also a function of the Feynman parameter $x$ implicitly.

The fermion loops are given as \cite{Choi:2023cqs}
\begin{equation}
\begin{split}
 -i \tilde \Sigma_1^{f \bar f}(p^2) &= (-1)\cdot  \left( -i \frac{m_f}{v} \right)^2 N_c \int \frac{d^4 k}{(2\pi)^4} \text{tr} \left[ \frac{i (\slashed{k} + m_f)}{k^2 - m_f^2 + i\epsilon} \frac{i (\slashed{k} + \slashed{p} + m_f)}{(k+p)^2 - m_f^2 + i\epsilon} \right] \\
 & = - \frac{  m_f^2 N_c}{4 \pi^2 v^2} \textstyle \Big[ R_2(p,m_f) - 2 \int_0^1 dx  \Delta_{m_f} (p^2) Q_0(p,m_f) \Big],
\end{split}
\end{equation}
where $N_c$ is the number of colors, 3 for quarks and 1 for leptons.

The $W$ boson loop is decomposed into the loops of the transverse $W$ boson, the charged Goldstone $\phi^\pm$, the mixed $W$--Goldstones and the charged ghosts $\bar c^\pm c^\mp$
\be \label{SigmaW}
 \tilde\Sigma_{1}^{W} (p^2) \equiv \tilde\Sigma_{1}^{W^+ W^-}  (p^2) + \tilde \Sigma_{1}^{\phi^+ \phi^-}(p^2) +\tilde \Sigma_{1}^{W^{\pm} \phi^{\mp}} (p^2) + \tilde \Sigma_{1}^{\bar c^\pm c^\mp} (p^2).
\ee
Here,
\begin{align} \label{transvW}
\begin{split}
 \tilde \Sigma_{1}^{W^+ W^-} (p^2) &=  i \int \frac{d^4 k}{(2\pi)^4}  \frac{2i g_{\mu\nu}  m_W^2}{v}  \frac{-i g^{\mu\rho}}{k^2 - m_W^2 + i\epsilon} \frac{-i g^{\nu\sigma}}{(k+p)^2 - m_W^2 + i\epsilon} \frac{2i g_{\rho\sigma}  m_W^2}{v}  \\
 &=  - \frac{m_W^4}{\pi^2v^2} R_0(p,m_W),
\end{split} \\
\begin{split}
 \tilde \Sigma_{1}^{\phi^+ \phi^-}  (p^2) &= i \left( -i \frac{m_h^2}{ v} \right)^2 \int \frac{d^4 k}{(2\pi)^4} \frac{i}{k^2 - m_W^2 + i\epsilon} \frac{i}{(k+p)^2 - m_W^2 + i\epsilon} \\
& = - \frac{ m_h^4}{16 \pi^2 v^2} R_0(p,m_W),
 \end{split} \\
\label{mixedW}
\begin{split}
  \tilde \Sigma_{1}^{W^{\pm} \phi^{\mp}}(p^2) &= 2i  \int \frac{d^4 k}{(2\pi)^4}   \frac{ \pm im_W(k- p)_\mu }{v }  \frac{- i g^{\mu\rho}}{k^2 - m_W^2 + i\epsilon} \frac{i}{(k+p)^2 - m_W^2 + i\epsilon} \frac{\mp im_W(p-k)_\rho }{v }  \\
 & = - \frac{ m_W^2}{8 \pi^2 v^2 }  \left[ R_2(p,m_W) - \textstyle \int_0^1 dx ((1+x)^2 p^2 +2 \Delta_{m_W} (p^2)) Q_0(p,m_W) \right],
 \end{split}
\\ \label{cpcm}
\begin{split}
 \tilde \Sigma_{1}^{\bar c^\pm c^\mp} (p^2)&= 2i \cdot(-1)  \left( -i \frac{m_W^2}{v} \right)^2 \int \frac{d^4 k}{(2\pi)^4} \frac{i}{k^2 - m_W^2 + i\epsilon} \frac{i}{(k+p)^2 - m_W^2 + i\epsilon} \\
&= \frac{m_W^4}{8 \pi^2 v^2} R_0(p,m_W) .
 \end{split}
\end{align}
The factor of 2 in (\ref{mixedW}) is from the same contributions from $W^+\phi^-$ and $W^-\phi^+$. Also, since $\bar c^+,  c^-, \bar c^- ,c^+$ are all independent, we have two combinations in (\ref{cpcm}). The total sum (\ref{SigmaW}) is gauge invariant.

We have similar decomposition of the $Z$ boson loop: the loops of the transverse $Z$ boson, the neutral Goldstone $\varphi$, the mixed $Z$--Goldstone and the neutral ghost
\be
 \tilde\Sigma_{1}^{Z} (p^2) \equiv \tilde\Sigma_{1}^{ZZ} (p^2) + \tilde \Sigma_{1}^{\varphi \varphi}  (p^2)+\tilde \Sigma_{1}^{Z \varphi}  (p^2)+ \tilde \Sigma_{1}^{\bar c c} (p^2),
\ee
where
\begin{align}
\begin{split}
 \tilde \Sigma_{1}^{ZZ} (p^2)  &= \frac{i}{2}  \int \frac{d^4 k}{(2\pi)^4} \frac{2 i  m_Z^2 g_{\mu\nu}}{v}  \frac{-i g^{\mu\rho}}{k^2 - m_Z^2 + i\epsilon} \frac{-i g^{\nu\sigma}}{(k+p)^2 - m_Z^2 + i\epsilon} \frac{2 i  m_Z^2 g_{\rho\sigma}}{v} \\
 & =- \frac{  m_Z^4}{2\pi^2v^2} R_0(p,m_Z),
 \end{split} \\
\begin{split}
 \tilde \Sigma_{1}^{\varphi \varphi} (p^2)&= \frac{i}{2} \left( -i \frac{m_h^2}{ v} \right)^2 \int \frac{d^4 k}{(2\pi)^4} \frac{i}{k^2 - m_Z^2 + i\epsilon} \frac{i}{(k+p)^2 - m_Z^2 + i\epsilon} \\
& =-\frac{ m_h^4}{32 \pi^2 v^2} R_0(p,m_Z),
 \end{split}
\\
 \label{mixedZ}
\begin{split}
  \tilde \Sigma_{1}^{Z\varphi} (p^2)&=i \int \frac{d^4 k}{(2\pi)^4} \frac{m_Z(p-k)_\mu}{v} \frac{-i g^{\mu\rho}}{k^2 - m_Z^2 + i\epsilon} \frac{i}{(k+p)^2 - m_Z^2 + i\epsilon}\frac{ m_Z (k-p)_\rho}{v}  \\
& = - \frac{  m_Z^2}{16 \pi^2 v^2}  \left[  R_2(p,m_Z) - \textstyle \int_0^1 dx ((1+x)^2 p^2 + 2\Delta_{m_Z} (p^2)) Q_0(p,m_Z) \right],
 \end{split}
 \\
\begin{split}
\tilde \Sigma_{1}^{\bar c c} (p^2)&= i \cdot (-1)   \left( -i \frac{m_Z^2}{ v} \right)^2 \int \frac{d^4 k}{(2\pi)^4} \frac{i}{k^2 - m_Z^2 + i\epsilon} \frac{i}{(k+p)^2 - m_Z^2 + i\epsilon} \\
&= \frac{m_Z^4}{16 \pi^2 v^2} R_0(p,m_Z) .
 \end{split}
\end{align}

Finally, we have the Higgs self-interaction loop
\begin{equation}
\begin{split}
\tilde \Sigma_{1}^{hh} (p^2)  &= \frac{i}{2} \left( -i \frac{3 m_h^2}{v} \right)^2 \int \frac{d^4 k}{(2\pi)^4} \frac{i}{k^2 - m_h^2 + i\epsilon} \frac{i}{(k+p)^2 - m_h^2 + i\epsilon} \\
 & = - \frac{9 m_h^4}{32  \pi^2  v^2} R_0(p,m_h).
 \end{split}
\end{equation}

The main result, the one-loop corrected Higgs mass squared that is renormalized as in (\ref{SlidingMass}), is drawn in Fig. \ref{figMain}. It is characterized by a peak followed by a drastic decrease, caused by the renormalized mass with a power-law running: for sufficiently high energy, the mass scales as $p^2$. For sufficiently small $\sqrt{p^2}$, the loop correction is small compared to the pole mass and the $(p^2-m_h^2)$ expansion of the propagator in (\ref{gamma2}) is more reliable than we expected. 

\begin{figure}[t]
\includegraphics[trim={2cm 0 0 0},clip,scale=0.5]{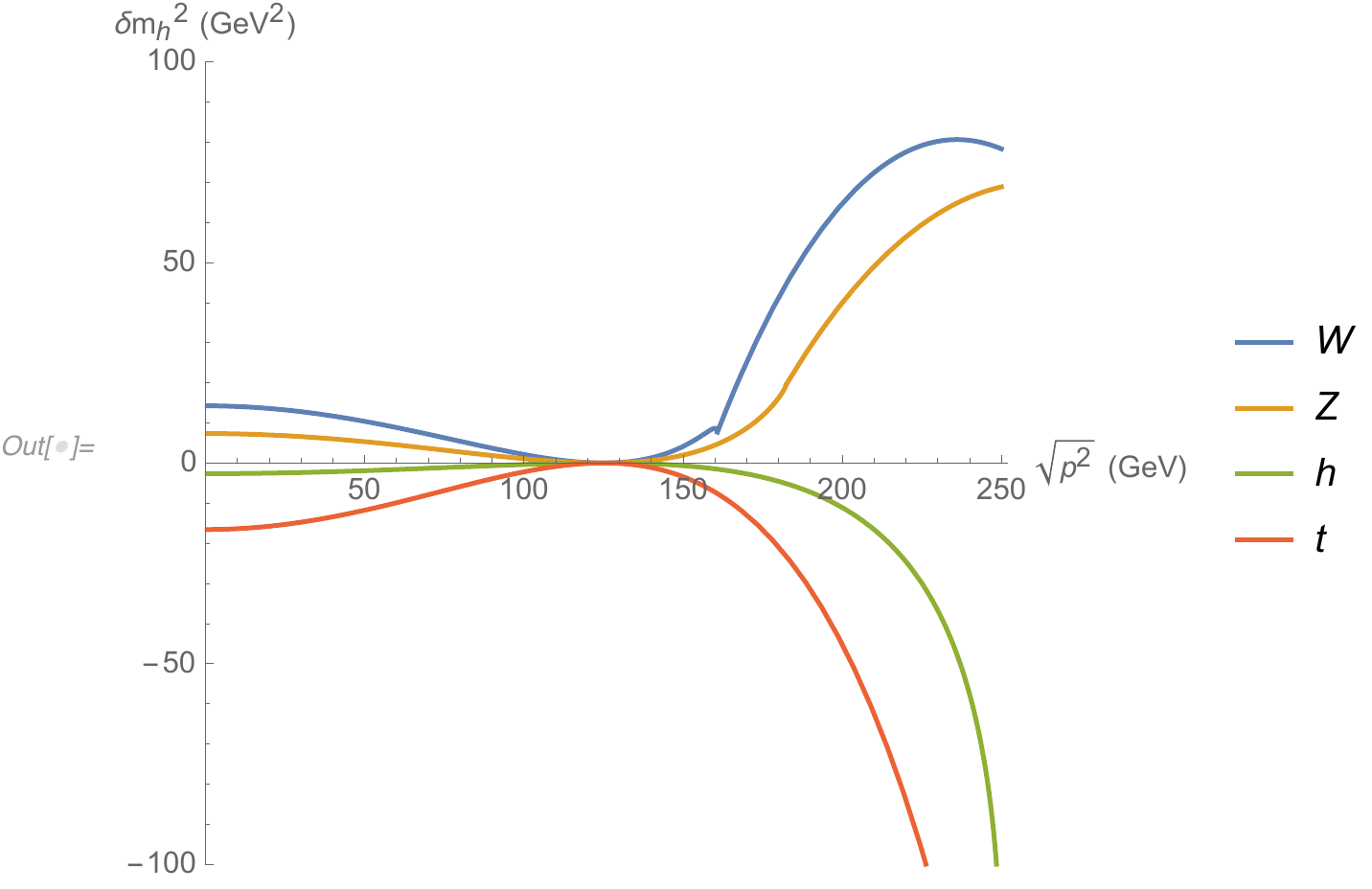} 
\caption{Individual contributions  from cubic interactions involving the $W$ boson, the $Z$ boson, the Higgs boson ($h$) and the top quark ($t$). These are the main contributions, while those from other fermions are negligibly small. We can see the cusps at $m_W$ and $m_Z$. Above them, the running becomes quadratic in the energy-momentum. \label{figContrib}}
\end{figure}

In Fig. \ref{figContrib}, we show the separate contributions from various fields. 
First, the largest contributions, slightly above $\sqrt{p^2} = m_h$, come from the $W$ boson loops. Then, the $Z$ boson contribution becomes sizable until $\sqrt{p^2}<2m_h$. We use the Feynman--'t Hooft gauge, in which the amplitudes of massive vector bosons have many components, as shown in Fig. \ref{figloops}. Among them, the mixed ones involving a vector boson and its associated Goldstone boson \eqref{mixedZ} and \eqref{mixedW} are particularly large, because the dimensionless couplings lead to amplitudes that scale quadratically with the internal momentum. For large external momentum, the top quark contribution dominates. The other fermion contributions are negligible. The remaining contributions, including the Higgs self-coupling loop, are logarithmic and subdominant.  

To see the cancellation of divergence, it is instructive to study the renormalized contribution from the transverse $W$ boson (\ref{transvW})
\be \begin{split}
  \tilde \Sigma_{1,\text{ren}}^{W^+ W^-} (p^2) &\equiv \tilde \Sigma_{1}^{W^+ W^-} (p^2)  -  \tilde \Sigma_{1}^{W^+ W^-} (m_h^2) - (p^2-m_h^2) \frac{ \tilde \Sigma_{1}^{W^+ W^-}}{dp^2} (m_h^2) \\
  &= - \frac{m_W^4}{\pi^2 v^2 } \int_0^1 dx  \left[ \log\frac{m_W^2-x(1-x)m_h^2}{m_W^2-x(1-x)p^2} -(p^2-m_h^2) \frac{x(1-x)}{m_W^2-x(1-x)m_h^2} \right] . 
\end{split}
\ee
We may verify that there is no logarithmic divergence, regardless of the choice of regularization. Although the superficial degree of divergence is zero, we need the $p^2$-derivative term from the construction \eqref{gamma2}
Even with the quadratic scaling of the self-energy, the correction with a similar definition,
\be \begin{split}
 \tilde \Sigma_{1,\text{ren}}^{W^{\pm} \phi^{\mp}}(p^2) = \frac{ m_W^2}{8 \pi^2 v^2 } \int_0^1 dx  &\left[ (2m_W^2+(3x^2+1)p^2) \log \frac{m_W^2-x(1-x)m_h^2}{m_W^2-x(1-x)p^2} \right. \\
 &\qquad \left. -(p^2-m_h^2) \frac{x(1-x)(2m_W^2+(3x^2+1)m_h^2)}{m_W^2-x(1-x)m_h^2} \right],
\end{split}
\ee
is free of quadratic divergence and the running behavior is essentially the same with enhanced power in $p^2$. 

The $W$-loop amplitudes have a threshold at $p^2 = 4 m^2_W$, opening the decay channel of the Higgs to two on-shell $W$ bosons. 
For $p^2 > 4 m^2_W$, the self-energy acquires an imaginary part and the mass correction is the real part of $\tilde \Sigma_{1,\text{ren}}^{W} (p^2)$ \cite{Hollik:1988ii}. Its running becomes approximately quadratic in $\sqrt{p^2}$. At the threshold, the mass squared typically has a kink. Below the threshold, the logarithm compensates for the power term, making the running mild. 
By the optical theorem, the imaginary parts of the self-energies correspond to the two-body decay widths $[\Im \tilde \Sigma^{W}_{1,\text{ren}}(p^2)]_ {p^2=m_h^2} = m_h \Gamma(h\to W^+ W^-)$ and  $[\Im \tilde \Sigma^{f \bar f}_{1,\text{ren}}(p^2)]_ {p^2=m_h^2} = m_h \Gamma(h\to f \bar f).$ We have verified these relations and found excellent agreement with the known results  \cite{Ellis:1975hr,Lee:1977eg,Gunion:1989we}. 

We call for experimental tests of the running Higgs mass. We may verify a deviation from the constant mass along the external momentum at the Large Hadron Collider. 
Within $p^2<4m_h^2$, the power-law running produces a local maximum peak with a deviation $48.40/15625 \simeq 0.31\%$ at $\sqrt{p^2}=199\,$GeV. The current observed mass is $125.11 \pm 0.11$ GeV (ATLAS) 
\cite{ATLAS:2023owm} and $125.04 \pm 0.12$ GeV (CMS) \cite{CMS:2024eka}, 
presumably at around the pole mass, which are converted to $15652.51 \pm 27.52 \,\text{GeV}^2$ and $15723.14 \pm 30.01 \,\text{GeV}^2$, respectively. These measurements have $0.17\%$ and $0.19\%$ precision, respectively, which is potentially sufficient to detect the predicted deviation, especially the increase beyond the pole mass $p^2 \gtrsim m_h^2$. If we can detect the subsequent peak, it would be definitive evidence.

Additionally, the observation of varying Higgs mass allows us to test the decoupling of heavy fields, which relies on this structure of renormalization. It makes the Higgs mass insensitive to ultraviolet physics and justifies the inclusion of only light fields in (\ref{totalSigma}) \cite{Choi:2024cbs, Choi:2024hkd}.

\begin{acknowledgements}
The author is grateful to Seungwon Baek and Tae-Jeong Kim for correspondence, and especially to Hyeseon Im for checking formulae.
This work is partly supported by the grant RS-2023-00277184 of the National Research Foundation of Korea.
\end{acknowledgements}

\end{document}